# Spin Coated Multilayered Cupric Oxide Thin Films and their structural properties

P. Samarasekara and N.G.K.V.M. Premasiri

Department of Physics, University of Peradeniya, Peradeniya, Sri Lanka

## Abstract

Sol-gel spin coating technique was employed to synthesize CuO thin films at 1500 rpm, 2200 rpm and 2400 rpm for 30 seconds, and subsequently annealed at 150-550°C for 1 hour. Multi-layered CuO thin films were prepared including 3 and 5 layers. These thin films were characterized using XRD, FTIR and polarizing light microscopy. The XRD and FTIR analysis confirmed the crystallization of CuO phase in thin films annealed above 350 °C, and corresponding crystallite sizes were observed to be increasing with the annealing temperature. The Scherrer formula was applied for (111) and ($\bar{1}$11) peaks of XRD patterns of CuO thin films. According to the data of ($\bar{1}$11) peak, average crystallite size increased from 0.801 to 55.2 $^0$A as annealing temperature was increased from 350 to 500 $^0$C.

## 1. Introduction

Although both cupric oxide and cuprous oxides are copper based compounds, cupric oxide possesses much desired characteristics in a host of applications. Cupric Oxide is a predominant candidate in a host of applications including photovoltaics [1], nanoelectronics [2] and spintronics [3]. CuO is a black color compound with slight translucence. This compound crystallizes in the monoclinic crystal system with lattice parameters: a = 4.684 Å, b = 3.425 Å, c = 5.129 Å and β = 99.28° [1]. Furthermore, this material possesses some unique antiferromagnetic properties, and recent studies have revealed superconductive properties (high temperature superconductivity) as well [4]. In the context of photovoltaics, CuO is an interesting semiconductor of p-type having a direct energy band gap of 1.2-1.9eV [1]. These characteristics distinguish CuO as a promising candidate in manifold applications, especially pertinent to solar energy. Moreover, it is important as a selective solar absorber due to its high solar absorbance and low thermal emittance.

When it comes to the realm of nanoelectronics, copper oxide is used increasingly in amultitude of applications mingling with cutting edge technology. For instance, copper oxide is used in the form of a thin film to fabricate copper oxide hetero structures; these structures are proven to be applicable for large-area electrical devices. CuO has been used in bilayer source-drain electrodes for organic thinfilm transistors (OTFTs) [2]. In addition, many different types of thin film transistors (TFT) are being experimented with this material. Cupric oxide has been proven to be a promising candidate in spintronics as well. CuO exhibits Room Temperature Ferromagnetism (RTF) [3], and thus this material offers a very good option for a class of spintronics, especially without the presence of any transition metal. Besides the preceding major applications, cupric oxide has been found indispensable for diode fabrication[Au/CuO/p-Si metal/interlayer/semiconductor (MIS)] [5], battery preparation (as electrodes for lithium-ion micro-batteries) [6], catalysis [7], electrochemical cells [8] and so on.

Different types of preparation techniques have been successfully implemented for cupric oxide: sol-gel spin coating [1, 9], dc sputtering [10], electrodeposition [11], spray pyrolysis technique [12], and so forth. In addition, photovoltaic [10] and gas sensing [13] properties of sputtered copper oxide films have been investigated. The methane gas sensing properties of cuprous oxide synthesized by thermal oxidation was studied [14]. Low Cost p-$Cu_2O$/n-CuO Junction has been characterized [15]. In addition, lithium mixed ferrite films have been grown using sputtering [16], and carbon nanotubes were synthesized using CVD method by us [17]. Copper oxide was fabricated using reactive dc sputtering by us [18]. Energy gap of semiconductor particles doped with salts were determined by us [19]. Cu based compounds indicate magnetic properties. The magnetic properties of ferromagnetic and ferrite thin films have been theoretically explained by us [20, 21, 22, 23, 24, 25].

In this report, structural properties of CuO thin films have been explained. Cupric acetate was used as the initial copper-containing material. In addition, two additives, namely ethylene glycol and polyethylene glycol were incorporated into the solution used for spin coating.

## 2. Experimental
Cupric acetate, diethanolamine and isopropyl alcohol were mixed together. The mole ratio between cupric acetate and diethanolamine was used to be 1:1. Then, a solution with $Cu^{+2}$ concentration of 1.5 mol $dm^{-3}$ was prepared, using isopropyl alcohol as the solvent. Then, this solution was continuously stirred for 24 hours. Thereafter, stirred solution was used for sol-gel spin coating. Finally, spin coated thin films were annealed. In addition to the basic solution of $Cu^{+2}$, 6 different solutions were prepared using those two additives; 3 solutions per each additive, varying w/w %. For this purpose, the additives were included in the solution at the initial step of the experimental procedure. Three different spin rates (1500 rpm, 2200 rpm, 2400 rpm), and two spin times (15 sec., 60 sec.) were utilized to characterize thin films. On top of that, five different annealing temperature values with an increment of 100 °C (150 °C, 250 °C, 350 °C, 450 °C, 550 °C), and two different annealing times (1 hour, 2 hour) were used in the annealing process.

Structural properties were determined using -ray diffractometer (SIEMENS D5000). In addition to the JCPDS data card for CuO, "XPowder12" software (Martillo Sigmar Software, Spain) was used in the analysis. This software is rather accurate and convenient, especially in being calculated Full Width at Half Maxima (FWHM:β) corresponding to different diffraction peaks. Fourier-transform infrared spectroscopy (FTIR) was determined using Shimadzu IRPrestige-21. Initially, a background analysis was performed using a cleaned piece of glass, and then thin film sample was analyzed for FTIR peaks. Universal Trinocular Polarizing Microscope (Euromex Universal Polarizing Microscope ME.2895) was employed to obtain the images of thin films.

## 3. Results and discussion

Two different ratios between cupric acetate and diethanolamine were experimented (1:1 and 1:2). Furthermore, three different concentrations for $Cu^{+2}$ were experimented (1mol $dm^{-3}$, 1.5 mol $dm^{-3}$, 2mol $dm^{-3}$). When $Cu^{+2}$ concentration was 1mol $dm^{-3}$, and the molar ratio was 1:2 (favoring DEA), thin film appeared to be patchy without having uniform distribution throughout the substrate, and this was observed in both as-prepared and after-annealed films prepared under those conditions [Figure 1 (a)]. Moreover, when $Cu^{+2}$ concentration was adjusted to be 2 mol $dm^{-3}$, film became uneven [Figure 1 (b)]. Consequently, it was found that 1:1 and 1.5 mol $dm^{-3}$ makes the most appropriate combination for those two parameters. Although, three different spinning rates (1500 rpm, 2200 rpm, 2400 rpm), and two spin times (15 sec., 60 sec.) were utilized, the appearance of the thin film did not show any differences at all.

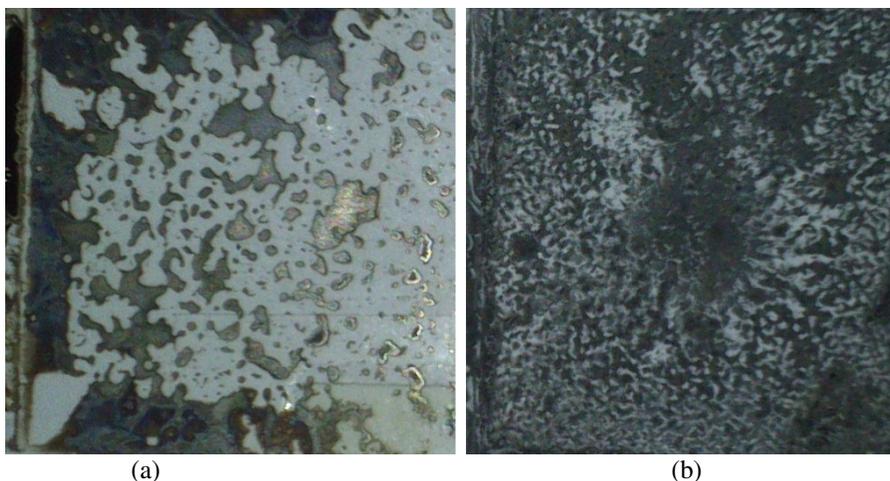

(a)                  (b)

*Fig. 1: (a) An annealed thin film prepared using a solution of $[Cu^{+2}]$ = 1 mol dm$^{-3}$ and $Cu^{+2}$: DEA = 1:2, (b) an annealed thin film prepared using a solution of $[Cu^{+2}]$ = 2 mol dm$^{-3}$ and $Cu^{+2}$: DEA = 1:2.*

Five different temperature values with an increment of 100 °C (150 °C, 250 °C, 350 °C, 450 °C, 550 °C), and two different annealing times (1 hour, 2 hour) were utilized. Thereupon, annealing time did not lead to any significant change in the appearance. But, annealing temperature was a very sensitive parameter in this aspect, because different annealing temperatures exhibited salient color changes in the thin films. Moreover, it was attempted to anneal spin-coated thin films above 550 °C, but they tended to be unstable, especially due to the deformations of the substrate.

In addition to those bulk appearances, annealed thin films were observed under a magnification of (100×) in order to find whether those thin films are adequately uniform in this exaggerated view as shown in figure 2.

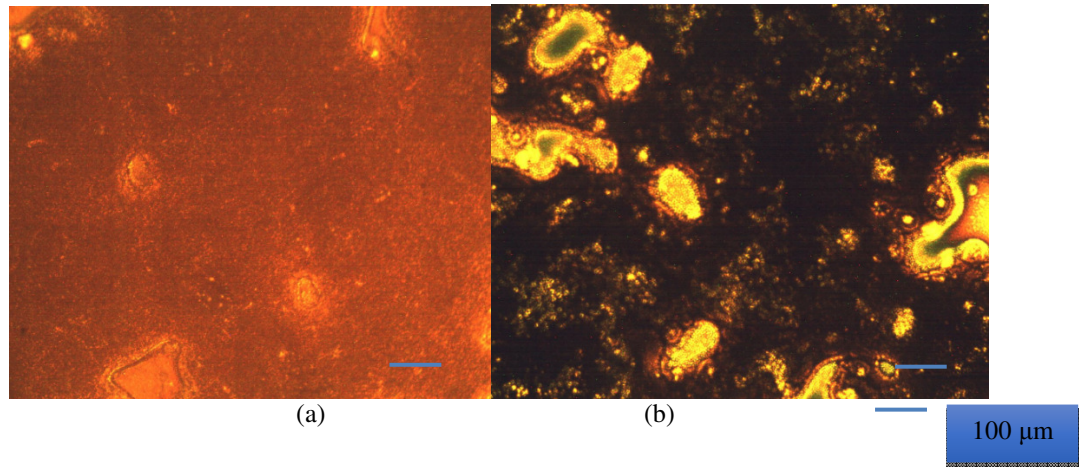

*Fig. 2: Magnified images of a thin film (annealed at 450 °C) obtained for the (a) reflection from the film, (b) transmission through the film; using a polarized light microscope.*

Figure 3 shows the X ray diffraction patterns (XRD) of a CuO thin film annealed at different temperatures. There are basically two diffraction peaks that can be used to verify the formation of crystalline CuO, in comparison with the JCPDS (Joint Committee on Powder Diffraction Standards) data card for CuO [Card No. 45-0937]. These two peaks indicate (111) and ($\bar{1}$11) lattice planes of Cu corresponding to standard lattice spacing (d values) of 2.3120 and 2.5230 $A^0$, respectively. Corresponding experimental d values are 2.3141 and 2.5130, respectively.

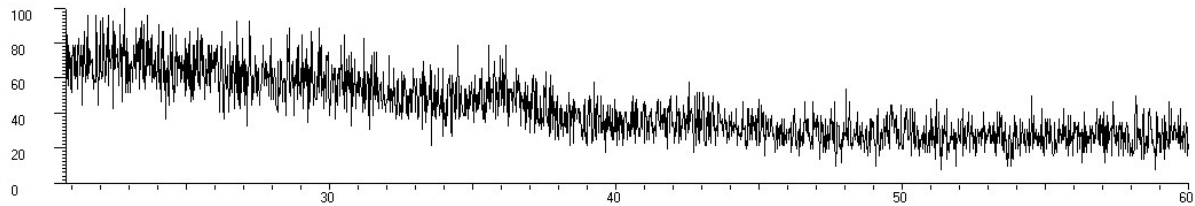

(a)

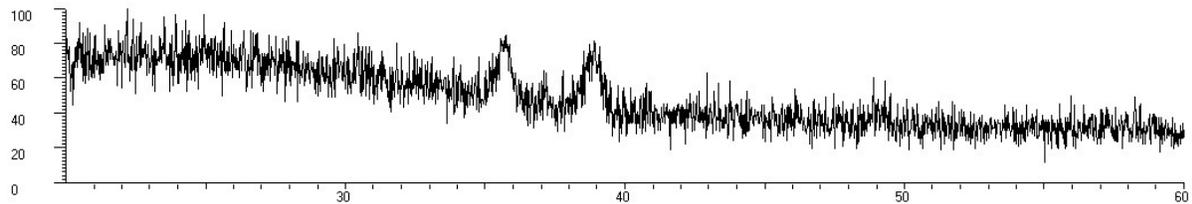

(b)

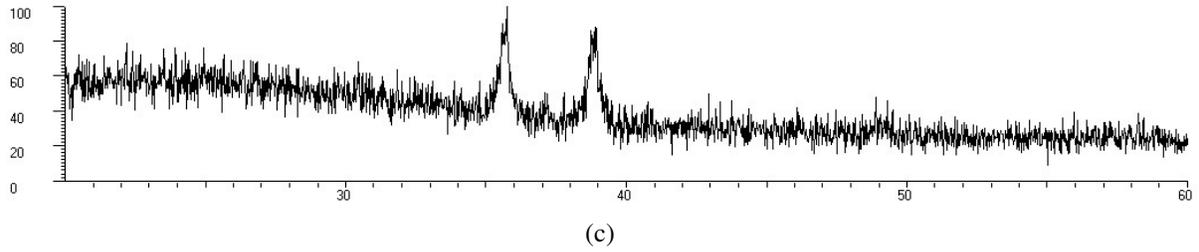

(c)

*Fig. 3: XRD pattern of a CuO thin film annealed at (a) 150 °C, (b) 350 °C and (c) 550 °C.*

The Scherrer equation, $L = \dfrac{\kappa \lambda}{\beta \cos\theta}$, was applied to approximate the crystallite size of the CuO crystals formed in the thin films. Here *L* is crystallite size, κ is geometrical factor related to the shape of the crystallites (For a spherical shape crystallite, κ ~ 1.07, and CuO crystallites can be assumed to be approximately spherical), λ is wavelength of the used X-rays, β is full width at half maximum (FWHM) of the diffraction peak and θ is diffraction angle. Crystallite sizes of the CuO crystals formed in spin-coated thin films are tabulated in table 1.

| Annealing Temperature (°C) | Crystal Plane | θ (°) | β (× $10^{-2}$ rad) | L (Å) |
| --- | --- | --- | --- | --- |
| 350 | ($\bar{1}$11) | 17.879 | 2.16 | 0.801 |
|  | (111) | 19.411 | 2.06 | 0.850 |
| 450 | ($\bar{1}$11) | 17.885 | 1.27 | 13.7 |
|  | (111) | 19.408 | 1.21 | 14.5 |
| 550 | ($\bar{1}$11) | 17.887 | 0.314 | 55.2 |
|  | (111) | 19.402 | 0.349 | 50.1 |

*Table 1: Crystallite sizes of samples annealed at different temperatures and spin rate = 2200 rpm; spin time = 60 sec.; annealing time = 1 hour, calculated using κ ~ 1.07 and λ = 1.5423 Å.*

Fourier-transform infrared spectroscopy can along with XRD analysis were used to confirm the formation of Cupric Oxide on the thin films. Thereupon, FTIR analyses were carried out for the five different temperature values that were used to anneal thin films.

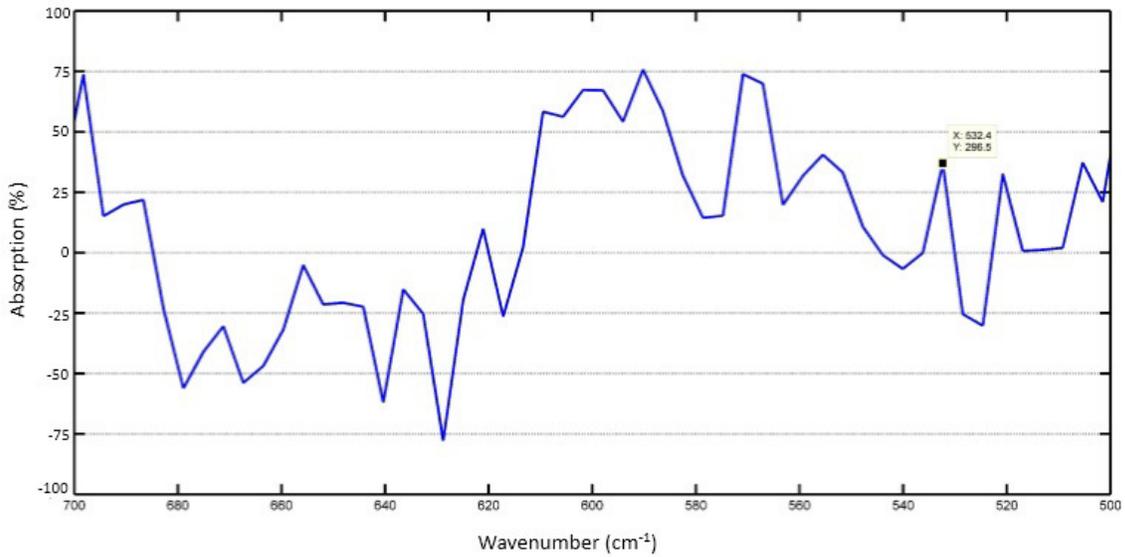

(a)

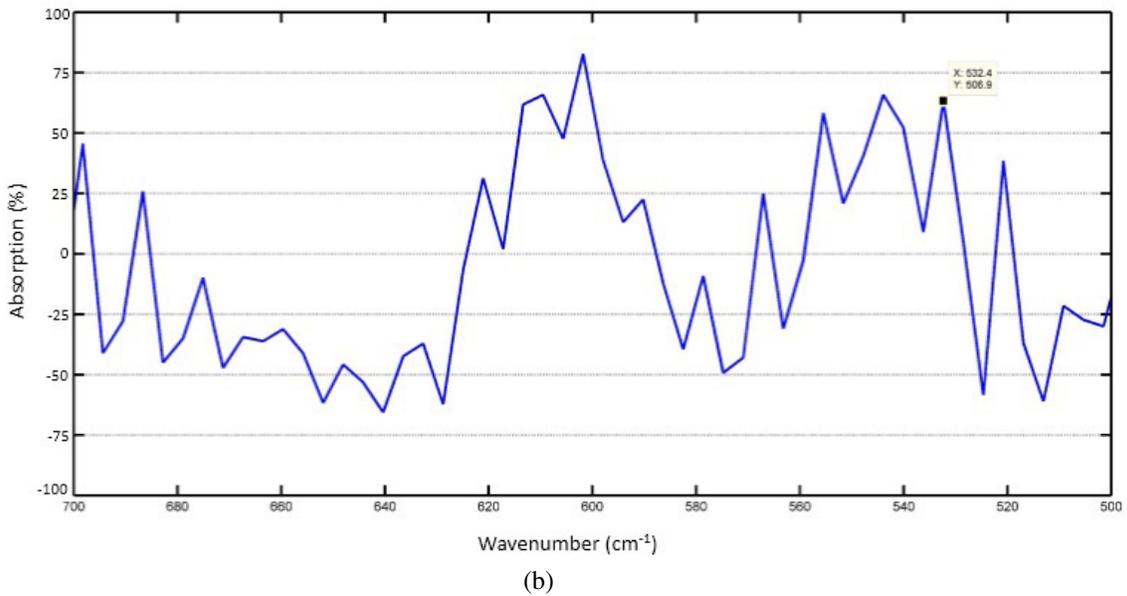

(b)

*Fig. 4: The FTIR spectrum of a CuO thin film annealed at (a) 350 and (b) 550 °C.*

The preceding five FTIR spectra provide an explicit trend with respect to the characteristic Cu-O bond stretching corresponding to the wave number 532 cm$^{-1}$ [1]. That is, as the annealing temperature is increased, the absorption percentage corresponding to the Cu-O bond stretching increases, implying the pronounced formation of cupric oxide at higher temperatures. Moreover, it can be observed that for the temperatures of 150 °C and 250 °C, Cu-O stretching does not exist at all, and the absorption begins to appear at 350 °C, and eventually at 350 °C and 550 °C this peak substantiates, convincing the formation of cupric oxide.

## 4. Conclusion

XRD analysis verifies the formation of cupric oxide in the spin-coated thin films. Furthermore, the strong concordance between the d values (distances between parallel atomic planes) corresponding to the XRD analyses which were done in this research work and the experimental standards (Joint Committee on Powder Diffraction Standards) elaborates on this conclusion. In addition, the sequence of XRD analyses in the figure 3 essentially conveys the appropriate annealing ambience required for a $Cu(OH)_2$ spin-coated thin film to convert that into (and crystallize) cupric oxide. This can be identified as annealing temperature of above 350 °C, but it is evident that annealing temperature becomes rather desirable when it is in the region of 450 - 550 °C, with increased intensities of diffraction peaks. Besides, approximating the crystallite size of the CuO crystals formed in the thin films using the Scherrer equation suggests that as the annealing temperature increases, corresponding crystallite size becomes larger. FTIR absorption peak of Cu-O bond appears only above 350 °C. Therefore, the FTIR spectrums confirm the crystallization of CuO phase in thin film above 350 $^0$C.

**References**

[1] D. Arun Kumar, Francis P. Xavier and J. Merline Shyla, 2012. Investigation on the variation of conductivity and photoconductivity of CuO thin films as a function of layers of coating. Archives of Applied Science Research 4 (5), 2174-2183.

[2] Jeong-Woo Park, Kang-Jun Baeg, Jieun Ghim, Seok-Ju Kang, Jeong-Ho Park and Dong-Yu Kim, 2007. Effects of copper oxide/gold electrode as the source-drain electrodes in organic thin film transistors. Electrochemical and Solid-State Letters 10 (11), 340-343.

[3] Daqiang Gao, Jing Zhang, Jingyi Zhu, Jing Qi, Zhaohui Zhang, Wenbo Sui, Huigang Shi and Desheng Xue, 2010. Vacancy-mediated magnetism in pure copper oxide nanoparticles. Nanoscale Research Letters 5, 769–772.

[4] Guanhua Chen, Jean-Marc Langlois, Yuejin Guo, and William A. Goddard, III, 1989. Superconducting properties of copper oxide high temperature superconductors. Proceedings of the National Academy of Sciences of U S A 86 (10), 3447–3451.

[5] İbrahim Y. Erdoğan, Ö. Güllü, 2010. Optical and structural properties of CuO nanofilms: Its diode application. Journal of Alloys and Compounds 492 (1–2), 378–383.

[6] E.A. Souza, R. Landers, L.P. Cardoso, Tersio G.S. Cruz, M.H. Tabacniks and A. Gorenstein, 2006. Evaluation of copper oxide thin films as electrodes for microbatteries. Journal of Power Sources 155 (2), 358-363.

[7] R. Manimaran, K. Palaniradja, N. Alagumurthi, S. Sendhilnathan and J. Hussain, 2014. Preparation and characterization of copper oxide nanofluid for heat transfer applications. Applied Nanoscience 4 (2), 163–167.

[8] L.B. Chen, N. Lu, C.M. Xu, H.C. Yu and T.H. Wang, 2009. Electrochemical performance of polycrystalline CuO nanowires as anode material for Li ion batteries. Electrochimica Acta 54 (17), 4198–4201.

[9] D.S.C. Halin, I.A. Talib, M. A. A. Hamid and A.R. Daud, 2008. Characterization of cuprous oxide thin films on n-Si substrate prepared by sol-gel spin coating. ECS Journal of Solid State Science and Technology 16 (1), 232-237.

[10] P. Samarasekara, M.A.K. Mallika Arachchi, A.S. Abeydeera, C.A.N. Fernando, A.S. Disanayake and R.M.G. Rajapakse, 2005. Photocurrent enhancement of D.C. Sputtered copper oxide thin films. Bulletin of Material Science 28(5), 483-486.

[11] T. Mahalingam, V. Dhanasekaran, G. Ravi, Soonil Lee, J. P. Chu and Han-Jo Lim, 2010. Effect of deposition potential on the physical properties of electrodeposited CuO thin films. Journal of optoelectronics and Advanced Materials 12 (6), 1327-1332.

[12] V. Saravanakannan and T. Radhakrishnan, 2014. Structural, electrical and optical characterization of CuO thin films prepared by spray pyrolysis technique. International Journal of ChemTech Research CODEN( USA) 6 (1), 306-310.

[13] P. Samarasekara, N.T.R.N. Kumara and N.U.S. Yapa, 2006. Sputtered Copper Oxide (CuO) thin films for Gas Sensor Devices. Journal of Physics Condensed Matter 18, 2417-2420.

[14] Ahalapitiya H. Jayatissa, P. Samarasekara and Guo Kun, 2009. Methane gas sensor application of cuprous oxide synthesized by thermal oxidation. Physica Status Solidi (a) 206 (2), 332-337.

[15] P. Samarasekara, 2010. Characterization of Low Cost p-$Cu_2$O/n-CuO Junction. Georgian Electronic Scientific Journals: Physics 2(4), 3-8.

[16] P. Samarasekara, 2002. Easy Axis Oriented Lithium Mixed Ferrite Films Deposited by the PLD Method. Chinese Journal of Physics 40(6), 631-636.

[17] P. Samarasekara, 2009. Hydrogen and Methane Gas Sensors Synthesis of Multi-Walled Carbon Nanotubes. Chinese Journal of Physics 47(3), 361-369.

[18] P. Samarasekara and N.U.S. Yapa, 2007. Effect of sputtering conditions on the gas sensitivity of Copper Oxide thin films. Sri Lankan Journal of Physics 8, 21-27.

[19] K. Tennakone, S.W.M.S. Wickramanayake, P. Samarasekara and, C.A.N. Fernando, 1987. Doping of Semiconductor Particles with Salts. Physica Status Solidi (a)104, K57-K60.

[20] P. Samarasekara, 2010. Determination of energy of thick spinel ferrite films using Heisenberg Hamiltonian with second order perturbation. Georgian electronic scientific journals: Physics 1(3), 46-49.

[21] P. Samarasekara, 2011. Investigation of Third Order Perturbed Heisenberg Hamiltonian of Thick Spinel Ferrite Films. Inventi Rapid: Algorithm Journal 2(1), 1-3.

[22] P. Samarasekara and William A. Mendoza, 2011. Third order perturbed Heisenberg Hamiltonian of spinel ferrite ultra-thin films. Georgian electronic scientific journals: Physics 1(5), 15-18.

[23] P. Samarasekara and William A. Mendoza, 2010. Effect of third order perturbation on Heisenberg Hamiltonian for non-oriented ultra-thin ferromagnetic films. Electronic Journal of Theoretical Physics 7(24), 197-210.

[24] P. Samarasekara, M.K. Abeyratne and S. Dehipawalage, 2009. Heisenberg Hamiltonian with Second Order Perturbation for Spinel Ferrite Thin Films. Electronic Journal of Theoretical Physics 6(20), 345-356.

[25] P. Samarasekara and Udara Saparamadu, 2013. Easy axis orientation of Barium hexa-ferrite films as explained by spin reorientation. Georgian electronic scientific journals: Physics 1(9), 10-15.